\documentstyle[aps,prb,epsf, multicol]{revtex}
\tighten
\begin{document}
\draft
\newcommand{\bn}{{\bf n}}
\newcommand{\bp}{{\bf p}}
\newcommand{\br}{{\bf r}}
\newcommand{\bq}{{\bf q}}
\newcommand{\bj}{{\bf j}}
\newcommand{\bE}{{\bf E}}
\newcommand{\eps}{\varepsilon}
\newcommand{\la}{\langle}
\newcommand{\ra}{\rangle}
\newcommand{\cK}{{\cal K}}
\newcommand{\cD}{{\cal D}}
\newcommand{\mybeginwide}{
    \end{multicols}\widetext
    \vspace*{-0.2truein}\noindent
    \hrulefill\hspace*{3.6truein}
}
\newcommand{\myendwide}{
    \hspace*{3.6truein}\noindent\hrulefill
    \begin{multicols}{2}\narrowtext\noindent
}

\title{
  Nonlocal effects in the shot noise of diffusive superconductor
  - normal-metal systems 
}
\author{K.\ E.\ Nagaev$^{1,2}$ 
}
\address{
  $^1$D\'epartement de Physique Th\'eorique, Universit\'e de Gen\`eve,
  CH-1211 Gen\`eve 4, Switzerland\\ 
  $^2$ Institute of Radioengineering and Electronics,
  Russian Academy of Sciences, Mokhovaya ulica 11, 103907 Moscow,
  Russia\\}
\date\today
\maketitle
\bigskip
\begin{abstract}
A cross-shaped diffusive system with two superconducting and two normal
electrodes is considered. A voltage $eV < \Delta$ is applied between the 
normal leads.
Even in the absence of average current through the superconducting electrodes 
their presence increases the shot noise at the normal electrodes and doubles 
it in the case of a strong coupling to the superconductors.
The nonequilibrium noise at the superconducting electrodes remains finite 
even in the case of a vanishingly small transport current due to
the absence of energy transfer into the superconductors. This noise is 
suppressed by electron-electron scattering at sufficiently high voltages.


\end{abstract}

\begin{multicols}{2}
\narrowtext


Recently, the noise properties of hybrid systems involving superconducting
(S) and normal (N) metals became a subject of intensive 
studies.\cite{Blanter-00} A key effect in these properties is the Andreev
reflection, in which electrons incident from the normal metal are reflected 
from the NS interface as holes. In particular, it was found that in the 
zero-voltage limit, the shot 
noise in diffusive NS contacts with phase-coherent transport is doubled with
respect to its value in a normal contact with the same 
resistance.\cite{deJong-94,Muzykantskii-94} This doubling was interpreted
as an effective doubling of electron charge and has been 
experimentally confirmed in a number of papers.\cite{Jehl-99,Jehl-00,%
Kozhevnikov-00}

Quite recently, it was shown that the doubled shot noise in diffusive NS 
contacts survives at finite voltages of the order of the energy 
gap.\cite{Nagaev-01a} Moreover, this noise does not require a phase 
coherence\cite{Belzig-01} and may be described in terms of a semiclassical 
Boltzmann - Langevin equation.\cite{Kogan-69} In this approach, the 
increased noise in NS systems is due to an excess heating of electron
gas rather than to the doubling of the effective charge.

Along with studying the shot noise in two-terminal systems, a considerable
attention received the noise in multiterminal structures. It was found that
the noise of normal-metal diffusive structures with more than two electrodes
may exhibit exchange effects\cite{Blanter-97}
and nonlocal effects.\cite{Sukhorukov-99} The latter imply that the noise 
in the system is affected by a presence of open contacts even in the absence 
of average current flow through them.

Several authors also calculated 
current correlations in single-channel multiterminal NS  
systems.\cite{Anantram-96,Lesovik-99,Torres-01} In particular, it was
found that the noise measured at the two normal electrodes in a four-terminal
system depends on the phase difference between the two superconducting 
electrodes.\cite{Anantram-96} A current noise was also 
calculated at a tip placed on a multimode quantum-coherent NS
structure.\cite{Gramespacher-00}

In this paper, we consider nonlocal effects in multiterminal  mesoscopic 
diffusive SN systems using the semiclassical description of noise. In 
particular, we report on a semiclassical "proximity" effect in the noise
where the current noise in a normal conductor is increased by its contact
with a superconductor. Another finding is that under certain conditions, 
a large noise may be induced in an SNS structure even by a small transverse
transport current.


Consider a cross-shaped contact with two superconducting and two normal-metal
electrodes (see Fig. 1). The normal-metal electrodes are kept at potentials 
$\pm V/2$, and the superconducting electrodes are kept at zero potential. It 
is also assumed 
that the cross is symmetric, i.e. it consists of two identical resistances $R$
connecting the crossing point with the normal electrodes and two identical 
resistances $r$ connecting it with the superconductors. Each resistor presents 
a long and narrow diffusive wire with the Thouless energy much smaller than the 
energy gap of the superconductor. The resistance of the crossing point is
assumed to be negligible. The applied voltage $eV$ 
\begin{figure}
 \vspace{-2mm}
 \centerline{
   \epsfxsize8cm
   \epsffile{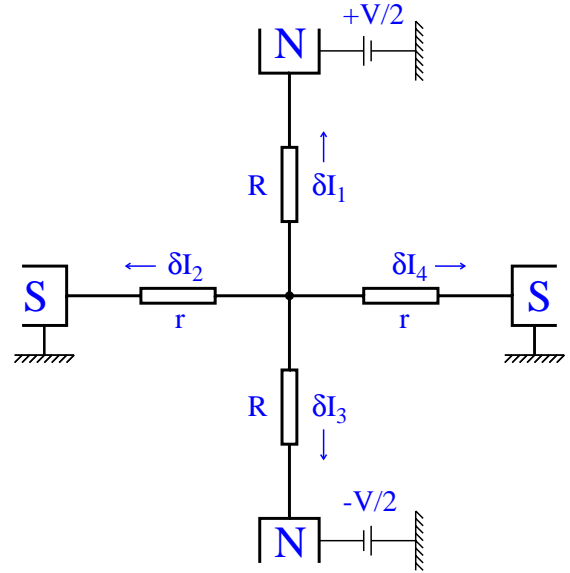}
 }
 \vspace{2mm}
 \caption{
   A four-terminal NS system. Normal-metal electrodes 1 and 3
   are kept at potentials $\pm V/2$, while the superconducting
   electrodes 2 and 4 are kept at zero potential.
 }
\label{FIG.1}
\end{figure}
\noindent
is assumed to be larger than the Thouless energy but 
smaller than the energy gap of the superconductors.
Note that in this geometry, there is now electrical current
flow through the superconducting ends of the cross.


Under the above conditions, the kinetics of fluctuations 
may be described using a semiclassical  Langevin 
equation\cite{Nagaev-01a}
\begin{equation}
 \delta\bj
 =
 -D
 \frac{ 
   \partial 
 }{ 
   \partial\br
 }
 \delta\rho
 -
 \sigma
 \frac{
   \partial
 }{
   \partial\br
 }
 \delta\phi
 +
 \delta\bj^{ext},
\label{dj}
\end{equation}
where $D$ is the diffusion coefficient, $\sigma$ is the electric
conductivity, $\delta\rho(\br)$ is the charge-density fluctuation,
$\delta\phi(\br)$ is the local fluctuation of the electric potential, and
the correlator of extraneous currents $\delta\bj^{ext}$ is given by
$$
 \la
  \delta j^{ext}_{\alpha}(\br_1)
  \delta j^{ext}_{\beta }(\br_2)
 \ra
 _{\omega}
 =
 4\sigma
 \delta_{\alpha\beta}
 \delta( \br_1 - \br_2 )
 T_N(\br_1),
$$
\begin{equation}
 T_N(\br)
 =
 \int d\eps\,
 f(\eps, \br)
 [
   1 - f(\eps, \br)
 ].
\label{T_N}
\end{equation}
Equation (\ref{dj}) may be integrated along the length of each arms
of the cross, which gives for the fluctuations of currents flowing 
into each of the four electrodes
\begin{equation}
 \delta I_i
 =
 \frac{1}{R_i}
 \left(
   \delta\phi^*
   +
   \frac{1}{e^2N_F}
   \delta\rho^*
 \right)
 +
 \delta I_i^{ext},
\label{dI_i=phi}
\end{equation}
where $R_i$ is the resistance of the corresponding arm, $\delta\phi^*$ 
and $\delta\rho^*$ are the fluctuations of electrical potential and charge
density at the crossing point, and the correlator of extraneous currents 
$\delta I_i^{ext}$ equals
\begin{equation}
 \la
  \delta I^{ext}_i(t_1)
  \delta I^{ext}_j(t_2)
 \ra
 _{\omega}
 =
 \delta_{ij}
 \frac{4T_i}{R_i},
\label{<dIdI>}
\end{equation}
where $T_i$ is obtained by averaging (\ref{T_N}) over the length of the 
corresponding arm. The system of equations (\ref{dI_i=phi}) combined with
the current-conservation condition at the crossing point
$$
 \sum\limits_i
 \delta I_i
 =
 0
$$
is easily solved giving
\begin{equation}
 \delta I_i
 =
 \delta I_i^{ext}
 -
 \frac{1}{R_i}
 \sum\limits_j
 \delta I_j^{ext}
 \left/
     \sum\limits_j
     \frac{1}{R_j}
 \right.,
\label{dI_i=sum}
\end{equation}
and hence the cross-correlated spectral density is
$$
 S_{ij}
 \equiv
 \la
   \delta I_i \delta I_j
 \ra_{\omega}
 =
 \delta_{ij}
 \frac{4T_i}{R_i}
 -
 \frac{4}{R_i R_j}
 \frac{
    T_i + T_j
 }{
    \sum\limits_k
    1/R_k
 }
$$
\begin{equation}
 +
 \frac{4}{R_i R_j}
 \sum\limits_k
 \frac{T_k}{R_k}
 \left/
    \left(
      \sum\limits_k
      \frac{1}{R_k}
    \right)^2
 \right..
\label{S_ij}
\end{equation}


In what follows, we restrict ourselves to relatively small
voltages $eV < 2\Delta$ and zero temperatures. In each arm of the cross, the 
average distribution function $f(\eps, x_i)$ obeys the standard diffusion 
equation $\nabla^2f = 0$. Introducing the distribution function $f^*(\eps)$ 
at the crossing point, one can write down the distribution in each arm in the 
form
\begin{equation}
 f(\eps, x_i)
 =
 \left(
   1 - \frac{x_i}{L_i}
 \right)
 f^*(\eps)
 +
 \frac{x_i}{L_i}
 f_i(\eps),
\label{f(eps,x_i)}
\end{equation}
where $L_i$ is the length of the arm and $f_i$ is the distribution at the 
end of it. The distribution function $f^*$ should be determined from
the balance of the diffusion fluxes at the crossing point for any energy
\begin{equation}
 \sum\limits_i
 \left.
   \frac{
     \partial f(\eps, x_i)
   }{
     \partial x_i
   }
 \right|_{x_i=0}
 =
 0.
\label{sum}
\end{equation}

The distribution functions at the normal ends of the cross $f_1 = f_0(\eps - 
eV/2)$ and $f_3 = f_0(\eps + eV/2)$ are just the equilibrium Fermi functions 
shifted in energy by $\pm eV/2$. Owing to the Andreev reflections from the 
interfaces with the superconductors, the  distribution functions at these 
interfaces are related to the distribution function at the crossing point by 
a very simple expression\cite{Nagaev-01a}
\begin{equation}
 f_2(\eps) 
 = 
 f_4(\eps)
 =
 \frac{1}{2}
 [
   1 + f^*(\eps) - f^*(-\eps)
 ].
\label{f_2}
\end{equation}
In the symmetric case where $R_1 = R_3 = R$ and $R_2 = R_4 = r$, it is easily 
obtained that
\begin{equation}
 f^*(\eps)
 = 
 f_2
 =
 f_4
 =
 \frac{1}{2}
 (
   f_1 + f_3
 ).
\label{f^*}
\end{equation}
%


Substituting Eq. (\ref{f^*}) into Eq. (\ref{f(eps,x_i)}), calculating the
corresponding $T_N(x_i)$ by means of (\ref{T_N}) and averaging them over the
corresponding segments, one easily obtains that $T_1 = T_3 = eV/6$ and 
$T_2 = T_4 = eV/4$. From this, one readily obtains the cross-correlated 
spectral densities by means of (\ref{S_ij}). Taking into account that the 
average current flowing between the normal electrodes is $I = eV/2R$, the 
expression for the noise at the normal ends may be written in the form
\begin{equation}
 S_{11}
 =
 S_{33}
 =
 \frac{eI}{3}
 \frac{
    4R^2 + 7Rr + 2r^2
 }{
    (R + r)^2
 },
\label{S_11}
\end{equation}
The cross-correlated noise at the normal ends is
\begin{equation}
 S_{13}
 =
 -
 \frac{eI}{3}
 \frac{
   r(R + 2r)
 }{
   (R + r)^2
 }.
\label{S_13}
\end{equation}
The spectral densities of noise at the superconducting ends of the cross
are given by
\begin{equation}
 S_{22}
 =
 S_{44}
 =
 \frac{eV}{6r}
 \frac{
   3R^2 + 8Rr + 6r^2
 }{
   (R + r)^2
 }
\label{S_22}
\end{equation}
and
\begin{equation}
 S_{24}
 =
 -
 \frac{eV}{6r}
 \frac{
   R(3R + 4r)
 }{
   (R + r)^2
 }.
\label{S_24}
\end{equation}
\begin{figure}
 \centerline{
   \epsfxsize8cm
   \epsffile{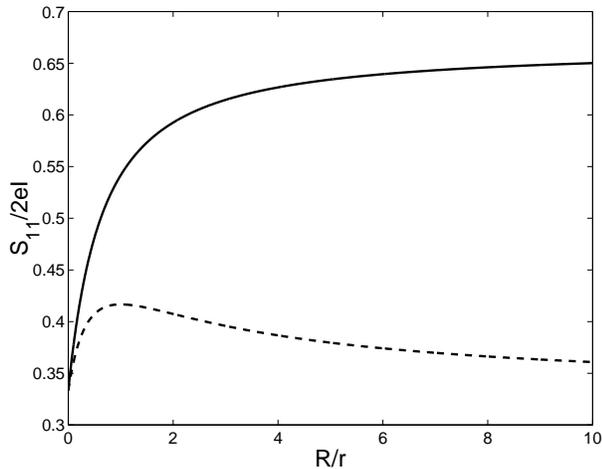}
 }
 \nopagebreak
 \vspace{3mm}
 \caption{
   The dependences of normalized spectral density of noise at 
   electrode 1 on the strength of coupling to the lateral
   electrodes 2 and 4 for a hybrid NS system (solid line) and
   for a normal-metal system (dashed line).
 }
\label{FIG.2}
\end{figure}
\noindent
To complete the description, we also present the cross-correlated spectral 
density at the normal and superconducting ends
\begin{equation}
 S_{12}
 =
 -
 \frac{eV}{6}
 \frac{
   3R + 2r
 }{
   (R + r)^2
 }.
\label{S_12}
\end{equation}
Note that all the cross-correlated spectral densities (\ref{S_13}), 
(\ref{S_24}), and (\ref{S_12}) are always negative.\cite{Nagaev-01a}


Consider now the most intresting case of a strong coupling to the 
superconductors where $r \ll R$. It is seen from Eqs. (\ref{S_11}) and
(\ref{S_13}) that the noise measured at one of the normal ends of the
cross is doubled with respect to a two-terminal normal-metal system, as it 
takes place in diffusive contacts where the transport current flows through
an SN interface\cite{deJong-94}. However the physics of this effect is
different because the cross-correlated spectral density of noise at the 
normal ends (\ref{S_13}) tends to zero. 

The reason for the increased shot noise is that the "superconducting"
arms, which do not carry any transport current, are open yet for the
current fluctuations and act as noise generators that supply additional
electric fluctuations into the system. In the case of normal-metal 
electrodes 2 and 4, the noise should be also slightly 
increased,\cite{Sukhorukov-99} but the increment would reach its 25\% 
maximum  ar $r = R$ (see Fig. 2). At strong coupling,
such cross would just break 
down into two independent contacts, each with the resistances $R$ and a 
voltage drop $V/2$. Hence the noise at electrodes 1 and 3 would be just
$2eI/3$. However the NS interfaces, while being transparent for the 
fluctuations of electric current, forbid the energy transport into the
superconductors and hence do not allow the distribution function at the
crossing point to assume the equilibrium shape. Clearly, the doubling
of shot noise should take place also in a three-terminal structure
with only one superconducting electrode attached in the middle of the
normal conductor.

It is noteworthy that the noise in the "superconducting" arms remains finite 
even if $R \to\infty$ and the transport current through the system vanishes. 
In this case, $S_{22} = S_{44} = -S_{24} = eV/2r$. This is in contrast to a
purely normal system, where $S_{22} = (2/3)eV/(R + r)$ and tends to zero as
either $r$ or $R$ becomes infinitely large. The reason for this is that the 
electron gas in this case is confined between two interfaces with  
superconductors, which hinder heat transfer from it. Hence it may be strongly 
heated even with a small current, much like as in the case of two-terminal 
diffusive SNS contacts it is strongly heated even by a small 
voltage\cite{Hoss-00,Nagaev-01b,Bezuglyi-01}. 

As $R\to 0$, the spectral densities
$S_{22}$ and $S_{44}$ increase to $eV/r$, while the cross-correlated spectral 
density $S_{24}$ tends to $-2eV/3r$, which implies that the current 
fluctuations at the opposite superconducting ends become only partially 
correlated. 


Consider now the effects of electron-electron scattering on the spectral
density $S_{22}$ in the case of a strong coupling. In the case of two-terminal 
SNS contacts, this type of scattering was shown to suppress the nonequilibrium
noise at low voltages\cite{Nagaev-01b,Bezuglyi-01} because the quasiparticles 
confined between the two NS interfaces may be outscattered from the subgap
energy range and escape into the superconducting electrodes thus transferring
energy and effectively cooling the electron gas.

Consider the case where
the length of a "normal" arm $L_R$ is much larger than the 
electron-electron scattering length. In this case, the electron gas may be
described by a local effective temperature, which depends only on the 
coordinate $x$ in the direction of average current flow and is constant in
the "superconducting" arms at $x = 0$. This effective temperature obeys a 
heat-balance equation\cite{Nagaev-95}
\begin{equation}
 \frac{\pi^2}{6}
 \frac{d^2}{dx^2}
 \left(
   T_e^2
 \right)
 =
 -
 \left(
   \frac{eV}{2L_R}
 \right)^2
 +
 \delta(x)
 \frac{2}{L_R\eps_T}
 \frac{R}{r}
 J,
\label{pi^2/6}
\end{equation}
where $\eps_T$ is the Thouless energy of a "superconducting" arm 
and 
$$
 J
 =
 \alpha_{ee}
 \frac{\Delta^3 T^*}{\eps_F}
 \exp
 \left(
    -\frac{\Delta}{T^*}
 \right)
$$
is the density of flux of energy carried by electrons and holes outscattered
from the subgap region by electron-electron collisions,\cite{Nagaev-01a}
which depends on the effective temperature in the "superconducting" arms
$T^* = T_e(0)$.
Making use of the boundary conditions $T_e(L_R) = T_e(-L_R) = 0$, one easily 
obtains a closed equation for $T^*$ in the form
\begin{equation}
  (T^*)^2
  =
  \frac{3}{4\pi^2}
  (eV)^2
  -
  \frac{6}{\pi^2}
  \alpha_{ee}
  \frac{R}{r}
  \frac{\Delta^3 T^*}{\eps_F\eps_T}
  \exp
  \left(
    -\frac{\Delta}{T^*}
  \right).
\label{T_e^2(0)}
\end{equation}
From this equation, it is readily seen that in the case  of large
values of the prefactor $\lambda = \alpha_{ee}(R/r)(\Delta^2/\eps_F\eps_T)$ 
the effective temperature $T^*$ and the spectral density $S_{22} = 2T^*/r$ 
become suppressed at voltages $V > \Delta/(e\ln\lambda)$. Unlike the case
of two-terminal SNS contacts, the suppression takes place at high voltages
rather than at small ones.


In summary, we semiclassically considered nonequilibrium noise in
diffusive multiterminal NS structures and found that nonlocal effects
in them are by far more pronounced than in purely normal systems with
the same geometry. In particular, they allow an observation of a doubled
longitudinal noise and of a giant transverse nonequilibrium noise with respect
to the direction of transport current.

The cross-shaped SN structures considered above are easily fabricated and 
hence the theoretical conclusions about the noise may be easily tested by
experimentalists. An advantage of this system is that there is no
voltage drop between the superconducting electrodes and the effects of electron
heating are not obscured by an onset of ac Josephson effect, as it
takes place for two-terminal SNS structures.\cite{Hoss-00}


I am grateful to W. Belzig for attracting my attention to multiterminal
geometries, to C. Strunk for a discussion of experimental results, and
to M. B\"uttiker for a discussion.

This work was supported by the Swiss National Science Foundation and 
Russian Foundation for Basic Research, Grant No. 01-02-17220.

\end{multicols}
\end{document}